\documentclass[preprint,review,12pt]{elsarticle}

\usepackage{amsmath}
\usepackage{amssymb}
\usepackage{amsfonts}
\usepackage{mathtools}
\usepackage{derivative}

\allowdisplaybreaks[1]

\DeclareMathOperator{\tr}{tr}
\DeclareMathOperator{\grad}{grad}
\DeclareMathOperator{\divg}{div}
\newcommand{\dd}[1]{\mathop{\mathrm{d}#1}}
\newcommand{\Obig}{\mathcal O}
\newcommand{\tens}[1]{\mathbf{#1}}

\newcommand{\EE}{\mathbf{E}}

\newcommand{\xs}{\tilde x}
\newcommand{\ts}{\tilde t}
\newcommand{\rs}{\tilde r}
\newcommand{\Us}{\tilde U}
\newcommand{\Vs}{\tilde V}

\newcommand{\us}{\tilde u}

\newcommand{\taus}{\tilde{\tau}}

\newcommand{\As}{\tilde{A}}

\newcommand{\ub}{\breve u}
\newcommand{\xib}{\breve\xi}
\newcommand{\taub}{\breve\tau}

\usepackage{graphicx}
\usepackage[usenames,dvipsnames,svgnames]{xcolor}
\usepackage{tikz}
\usetikzlibrary{
	calc,patterns,decorations.pathmorphing,decorations.markings,positioning}

\usepackage[top=20mm,bottom=20mm,left=20mm,right=20mm]{geometry}

\usepackage[unicode]{hyperref}
\hypersetup{
colorlinks=true,
linkcolor={black!50!black},
citecolor={blue!50!black},
urlcolor={blue!50!black},
}

\journal{International Journal of Non-Linear Mechanics}
\date{December 24, 2024}

\begin{document}

\begin{frontmatter}

\title{Slowly decaying strain solitons in nonlinear viscoelastic waveguides}

\author[1]{F.\,E. Garbuzov\corref{cor1}}
\ead{fedor.garbuzov@mail.ioffe.ru}
\cortext[cor1]{Corresponding author}

\author[1]{Y.\,M. Beltukov}

\affiliation[1]{organization={Ioffe Institute},
             addressline={26 Polytekhnicheskaya},
             city={St.~Petersburg},
             state={194021},
             country={Russia}}

\begin{abstract}
This paper is devoted to the modeling of longitudinal strain waves in a rod composed of a nonlinear viscoelastic material characterized by frequency-dependent second- and third-order elastic constants. 
We demonstrate that long waves in such a material can be effectively described by a damped Boussinesq-type equation for the longitudinal strain, incorporating dissipation through retarded operators.
Using the existing theory of solitary wave solutions in nearly integrable systems, we derive a slowly-decaying strain soliton solution to this equation.
The derived soliton characteristics are shown to be in a good agreement with results from full 3D simulations.
We demonstrate the importance of taking into account the frequency dependence of third-order elastic constants for the description of strain solitons.
\end{abstract}


\end{frontmatter}

\section{Introduction}
The study of solitonic structures and localized waves continues to attract the researcher's attention~\cite{Ostrovsky2024, KhusnutdinovaEditorial2022}. One of the ongoing studies is devoted to the strain solitary waves (solitons) in structural elements or waveguides such as bars, plates and shells~\cite{SamsonovBook, ShvartzAIP2015a, GKS2019}. This research extends to microstructured materials and acoustic metamaterials~\cite{Janno2005, Casasso2010, BerezovskiVan2017, Gula2018, Porubov2021}, where the nonlinear and dispersive properties that balance to form solitons can be even more pronounced.
Additionally, solitons are being explored for potential applications in nondestructive evaluation (NDE) tasks~\cite{DelsantoBook2006, KhusnSamsPhysRev2008, DreidenJAP2012, Khusn2017, Tranter2023}, which is complementary to the current development of nonlinear NDE techniques~\cite{Lissenden2021, Maev2022, Broda2014review}.

Theoretical models of strain solitons and nonlinear strain waves often assume the waveguide material to be purely elastic or, at most, incorporate simple linear viscoelasticity~\cite{Nariboli1970, Zabolotskaya2004, KhusnPRE2021, WaMot2022}.
However, recent studies have shown that certain polymers and soft materials exhibit strongly nonlinear viscoelastic properties, characterized by the pronounced frequency dependence of their third-order elastic constants (TOEC)~\cite{Chintada2022, TechPhys2021, Belashov2024}. For instance, in polystyrene low-frequency TOEC values turned out to be an order of magnitude larger in absolute value than high-frequency values~\cite{TechPhys2021, Belashov2024}. To model these effects, we developed a second-order nonlinear generalization of the standard linear solid within the framework of multiple-integral approach, providing an adequate description of the experimental observations~\cite{IJNLM2024}.

The purpose of the present study is to comprehensively extend the theory of strain solitons in elastic waveguides to nonlinear viscoelastic waveguides, and also to show how the general viscoelastic model, which we proposed in Ref.~\cite{IJNLM2024}, can be applied for the modeling of strain waves.
This work complements other theoretical studies on strain waves in nonlinear viscoelastic solids~\cite{DePascalis2019, Favrie2023}, motivated by reported nonlinearity and viscoelasticity of geomaterials~\cite{Martin2019, Tutuncu1998} and soft biological tisues~\cite{Darvish2001, Benitez2017}. 

The paper is structured as follows. Section 2 presents the general three-dimensional equations of motion for isotropic nonlinear viscoelastic solids. Section 3 derives a single wave equation from the general 3D framework, governing longitudinal strain waves in thin viscoelastic waveguides. The decaying solitary wave solutions of this equation are studied in Sec.~4. In Section 5, the derived theory is compared against the results of numerical simulations of the full 3D equations of motion. The paper concludes with a summary of findings in Section 6 and additional technical details provided in the Appendix.

\section{General equations}\label{sec:general}

The equations of motion of a deformable body and the free surface boundary conditions have the following form:
\begin{align}
\label{eq:motion}
	&\rho \ddot{\vec U} = \divg \tens P, &&\vec x \in \Omega,\\
\label{eq:bc}
	&\tens P \cdot \vec n = 0, &&\vec x \in \partial\Omega,
\end{align}
where $\vec{x}$ represents the material coordinates, $\rho$ is the material density, and $\vec{U}$ is the displacement vector. The top dot denotes a time derivative, while the central dot represents an inner product. The vector $\vec{n}$ is the unit normal to the body surface, $\tens{P}$ is the first Piola-Kirchhoff stress tensor, and the divergence of a tensor is defined as $(\divg \tens P)_i = \sum_j \partial_j P_{ij}$.

Tensor $\tens P$ is connected to the second Piola-Kirchhoff stress tensor $\tens{S}$ via the deformation gradient as:
\begin{equation} \label{eq:relation_pk1_pk2} 
	\tens{P} = (\tens{I} + \grad \vec{U}) \cdot \tens{S},
\end{equation}
where $\tens{I}$ is the identity tensor, and $\grad \vec{U}$ represents the displacement gradient, whose elements are of the form $(\grad \vec{U})_{ij} = \partial_j U_i$. 

\subsection{General constitutive equation}

The current state of a viscoelastic material is inherently dependent on its strain history. To capture this behavior, internal state variables are introduced, each of which describes some relaxation process occurring during deformation. The constitutive model applied in this work is based on power series expansion of stress tensor in the internal state variables~\cite{IJNLM2024, ChristensenBook, FindleyBook}:
\begin{equation} \label{eq:S_nonlin_internal_strains}
	\mathbf{S} = \sum_{s=0}^q \tens{C}_{s} : \EE_{s}  
	+ \sum_{s=0}^q\sum_{u=0}^q \tens{N}_{su} :: \EE_{s} \EE_{u}.
\end{equation}
Here, $\EE_s$ are the internal strain tensors (internal state variables), $\tens C_s$ are the fourth order tensors of linear viscoelastic moduli, $\tens N_{su}$ are the sixth order tensors of nonlinear viscoelastic moduli. The operators $:$ and $::$ denote double and quadruple contractions, respectively. In the coordinate form, these contractions are defined as ${(\mathbf{A}:\mathbf{B})}_{ij} = \sum_{kl}A_{ijkl}B_{kl}$ and ${(\mathbf{F}::\mathbf{B}\mathbf{D})}_{ij} = \sum_{klmn}F_{ijklmn}B_{kl}D_{mn}$, where $\mathbf A$, $\mathbf B$, $\mathbf D$, and $\mathbf F$ are arbitrary tensors.

The internal strains, \(E_s\), encapsulate the relaxation processes that occur during deformation and are subject to the following evolution equations:
\begin{equation}\label{eq:internal_strain_lin}
	\dot{\EE}_{s} + \frac{\EE_{s}}{\tau_s} = \dot{\EE}, \quad
	s = 1, \dots q,
\end{equation}
where $\tau_s$ is the characteristic relaxation time of the $s$-th process and $\EE$ is the Green-Lagrange finite strain tensor defined as follows:
\begin{equation}\label{eq:strain}
	\EE = \frac12 \Bigl[
		\grad \vec{U} 
		+ \bigl(\grad \vec{U}\bigr)^T 
		+ \bigl(\grad \vec{U}\bigr)^T \cdot \grad \vec{U}
	\Bigr].
\end{equation} 
Integration of Eq.~\eqref{eq:internal_strain_lin} provides the explicit dependence of the internal strains on the strain history:
\begin{equation}\label{eq:internal_strain_lin_int}
	\EE_s = \int_{-\infty}^{t} 
		e^{-\frac{t-t_1}{\tau_s}} \dot \EE(t_1) \dd{t_1}.
\end{equation}

As follows from Eq.~\eqref{eq:internal_strain_lin}, there are $q$ relaxation processes in the model.
The process with number $s$ is responsible for attenuation of harmonics with frequencies close to $(2\pi\tau_s)^{-1}$ and the number of relaxation processes to include in the model is defined by the range of frequencies that needs to be covered~\cite{Carcione2014}.
If certain processes have significantly larger relaxation times than the modeling timescale, they can be incorporated into the model as quasi-static elastic properties corresponding to the infinite relaxation time:
\begin{equation} \label{eq:tau0}
	\tau_0 = \infty, \quad 
	\EE_0 = \EE.
\end{equation}

\begin{figure}[h]
\centering
\begin{tikzpicture}[scale=0.8]
	\tikzstyle{spring1}=[thick,decorate,decoration={coil,
		aspect=0.5, 
		segment length=2mm, 
		amplitude=2mm, 
		pre length=3mm,
		post length=3mm}]
	\tikzstyle{damper1}=[thick,decoration={markings,  
		mark connection node=dmp,
		mark=at position 0.5 with 
		{
			\node (dmp) [thick,inner sep=0pt,transform shape,rotate=90,minimum width=12pt,minimum height=3pt,draw=none] {};
			\draw[thick] ($(dmp.north east)+(-3pt,0)$) -- (dmp.south east) -- (dmp.south west) -- ($(dmp.north west)+(-3pt,0)$);
			\draw[thick] ($(dmp.north)+(0,-4.5pt)$) -- ($(dmp.north)+(0,4.5pt)$);
	}}, decorate]
	\newcommand{\LenGenMax}{5}
	\newcommand{\GenMaxStep}{\LenGenMax/8}
	
	\draw[thick,stealth-] (0,\LenGenMax/2) -- ++(\GenMaxStep,0);
	\draw[thick] (\GenMaxStep,0) -- ++(0,\LenGenMax/8);
	\draw[thick,dashed] (\GenMaxStep,\LenGenMax/8) -- ++(0,\LenGenMax/4);
	\draw[thick] (\GenMaxStep,3*\LenGenMax/8) -- ++(0,5*\LenGenMax/8);
	\draw[spring1] (\GenMaxStep,\LenGenMax) -- ++(6*\GenMaxStep,0) node [midway, above=1.2mm] {$C_0$};
	\draw[spring1] (\GenMaxStep,3*\LenGenMax/4) -- ++(4*\GenMaxStep,0) node [midway, above=1.2mm] {$C_1$}; 
	\draw[damper1] (5*\GenMaxStep,3*\LenGenMax/4) -- ++(2*\GenMaxStep,0) node [midway, above=1.2mm] {$\tau_1$};
	\draw[spring1] (\GenMaxStep,\LenGenMax/2) -- ++(4*\GenMaxStep,0) node [midway, above=1.2mm] {$C_2$}; 
	\draw[damper1] (5*\GenMaxStep,\LenGenMax/2) -- ++(2*\GenMaxStep,0) node [midway, above=1.2mm] {$\tau_2$};
	\node at (4*\GenMaxStep,\LenGenMax/4) {{\Large\dots}};
	\draw[spring1] (\GenMaxStep,0) -- ++(4*\GenMaxStep,0) node [midway, above=1.2mm] {$C_q$}; 
	\draw[damper1] (5*\GenMaxStep,0) -- ++(2*\GenMaxStep,0) node [midway, above=1.2mm] {$\tau_q$};
	\draw[thick] (7*\GenMaxStep,0) -- ++(0,\LenGenMax/8);
	\draw[thick,dashed] (7*\GenMaxStep,\LenGenMax/8) -- ++(0,\LenGenMax/4);
	\draw[thick] (7*\GenMaxStep,3*\LenGenMax/8) -- ++(0,5*\LenGenMax/8);
	\draw[thick,-stealth] (7*\GenMaxStep,\LenGenMax/2) -- ++(\GenMaxStep,0);
\end{tikzpicture}
\caption{Schematic of the generalized standard linear solid. Variables $C_s$ denote the spring stiffnesses and $\tau_s$ represent the spring-dashpot relaxation times.}
\label{fig:gen_sls}
\end{figure}
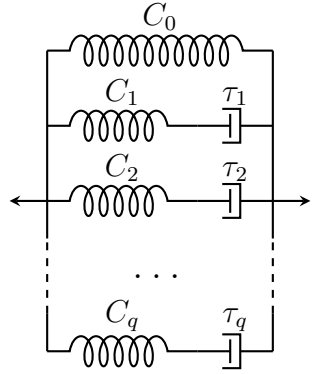

The described model represents the nonlinear generalization of the generalized standard linear solid. The latter, in the case of one-dimensional deformations, is usually illustrated by the spring-dashpot system in Fig.~\ref{fig:gen_sls}.
The nonlinear model described by Eqs.~\eqref{eq:S_nonlin_internal_strains} and \eqref{eq:internal_strain_lin} captures the nonlinear influence of relaxation processes on the total stress. As highlighted in Ref.~\cite{IJNLM2024}, an additional nonlinear effect may arise from the coupling between relaxation processes. However, this effect is not considered in the present study and its implications remain a subject for future research.

\subsection{Isotropic material}

In this paper, we consider an isotropic material, thus the elements of the fourth-order linear viscoelastic tensor $\tens C_s$ are defined only by two independent moduli $\lambda_s$ and $\mu_s$, and the elements of the sixth-order nonlinear viscoelastic tensor $\tens N_{su}$ can be expressed, in general, via the four moduli. However, as we discussed in Ref.~\cite{IJNLM2024}, it could be a challenging problem to estimate all four moduli independently from the experimental data, and it is much easier to measure only the three of them. Therefore, we assume that an isotropic tensor $\tens N_{su}$ is defined by only the three moduli $l_{su}$, $m_{su}$, and $n_{su}$, so that the viscoelastic tensors take the following form:
\begin{align}
	(C_s)_{ijkl} =&\, \lambda_s \delta_{ij}\delta_{kl} 
		+ 2\mu_s \mathcal{I}_{ijkl}, 
\label{eq:Cs_isotrop}\\
	(N_{su})_{ijklmn} =& \, \left(
		l_{su} - m_{su} + \frac{n_{su}}2 \right) 
		\delta_{ij}\delta_{kl}\delta_{mn} 
	+ \left(m_{su} - \frac{n_{su}}2\right) \left(
    	\delta_{ij} \mathcal{I}_{klmn} +  
    	\delta_{kl} \mathcal{I}_{ijmn} +
    	\delta_{mn} \mathcal{I}_{ijkl} \right) \nonumber\\
	& + \frac{n_{su}}4 \big(
	\delta_{ik}\mathcal{I}_{jlmn} +
	\delta_{jk}\mathcal{I}_{ilmn} +
	\delta_{il}\mathcal{I}_{jkmn} +
	\delta_{jl}\mathcal{I}_{ikmn}\big),
\label{eq:Nsu_isotrop}
\end{align}
where $\delta_{ij}$ denotes the Kronecker delta, $\mathcal I_{ijkl} = \frac12(\delta_{ik}\delta_{jl} + \delta_{il}\delta_{jk})$, and the sets of viscoelastic moduli posses the following properties:
\begin{gather}
\label{eq:lmn_symm}
	l_{su} = l_{us}, \quad m_{su} = m_{us}, \quad n_{su} = n_{us}.
\end{gather}
More on this can be found in Ref.~\cite{IJNLM2024}. We leave the investigation of the case of four moduli in the nonlinear viscoelastic tensor for future work.

\newcommand{\lamop}{\hat{\lambda}}
\newcommand{\muop}{\hat{\mu}}
\newcommand{\lop}{\hat{l}}
\newcommand{\mop}{\hat{m}}
\newcommand{\nop}{\hat{n}}
\newcommand{\Rop}{\hat{\mathcal{R}}}
Under the above assumptions, the stress-strain relation in an isotropic material can be expressed as  follows:
\begin{align}
\tens{S} = \tens I \lamop[\tr\EE] + 2\muop[\EE] 
	+ \tens I \left(\lop - \mop + \frac\nop 2 \right) 
	[\tr\EE, \tr\EE]
	+ \tens I \tr\left(\biggl(\mop - \frac\nop 2\biggr) [\EE,\EE]\right) \nonumber\\
	+ 2\left(\mop - \frac\nop 2 \right)[\EE,\tr\EE] + 
	\nop[\EE,\EE],
\label{eq:S_isotrop_op}
\end{align}
where $\tens I$ is the identity tensor, $\lamop$ and $\muop$ are the linear retarded integral operators, and $\lop$, $\mop$, and $\nop$ are the similar bilinear operators. These operators act on arbitrary functions $f$ and $g$ as follows:
\begin{align}
	\lamop[f] = \sum_{s=0}^q& \lambda_s \Rop_s[f], \quad
	\muop[f] = \sum_{s=0}^q \mu_s \Rop_s[f],  
\label{eq:lame_op}\\
	\lop[f, g] &= \sum_{s=0}^q\sum_{u=0}^q 
		l_{su} \Rop_s[f] \cdot \Rop_u[g],
\end{align}
operators $\mop$ and $\nop$ are defined similarly, 
and the definition of the retarded operators $\Rop_s$ follows from Eq.~\eqref{eq:internal_strain_lin_int}:
\begin{equation}\label{eq:R_op}
	\Rop_s[f] = \begin{dcases}
		\int\limits_{-\infty}^{t} e^{-\frac{t-t_1}{\tau_s}} 
			\dot f(t_1) \dd{t_1}, & s \geqslant 1,\\
		f, & s = 0,
	\end{dcases}
\end{equation}
with $\Rop_0$ being simply the identity operator which helps to take into account the quasi-static elastic properties.

We notice that if no relaxation process occurs in the material ($q = 0$) then Eq.~\eqref{eq:S_isotrop_op} describes an absolutely elastic isotropic material, the properties of which are characterized by the two Lame elastic moduli $\lambda_0$ and $\mu_0$ and the three Murnaghan moduli $l_{00}$, $m_{00}$, and $n_{00}$. This model was typically used in the previous studies on solitons in isotropic elastic waveguides~\cite{SamsonovBook, GKS2019}.

\section{Longitudinal waves in thin waveguides}
\begin{figure}[h]
\centering
\includegraphics[width=.4\textwidth]{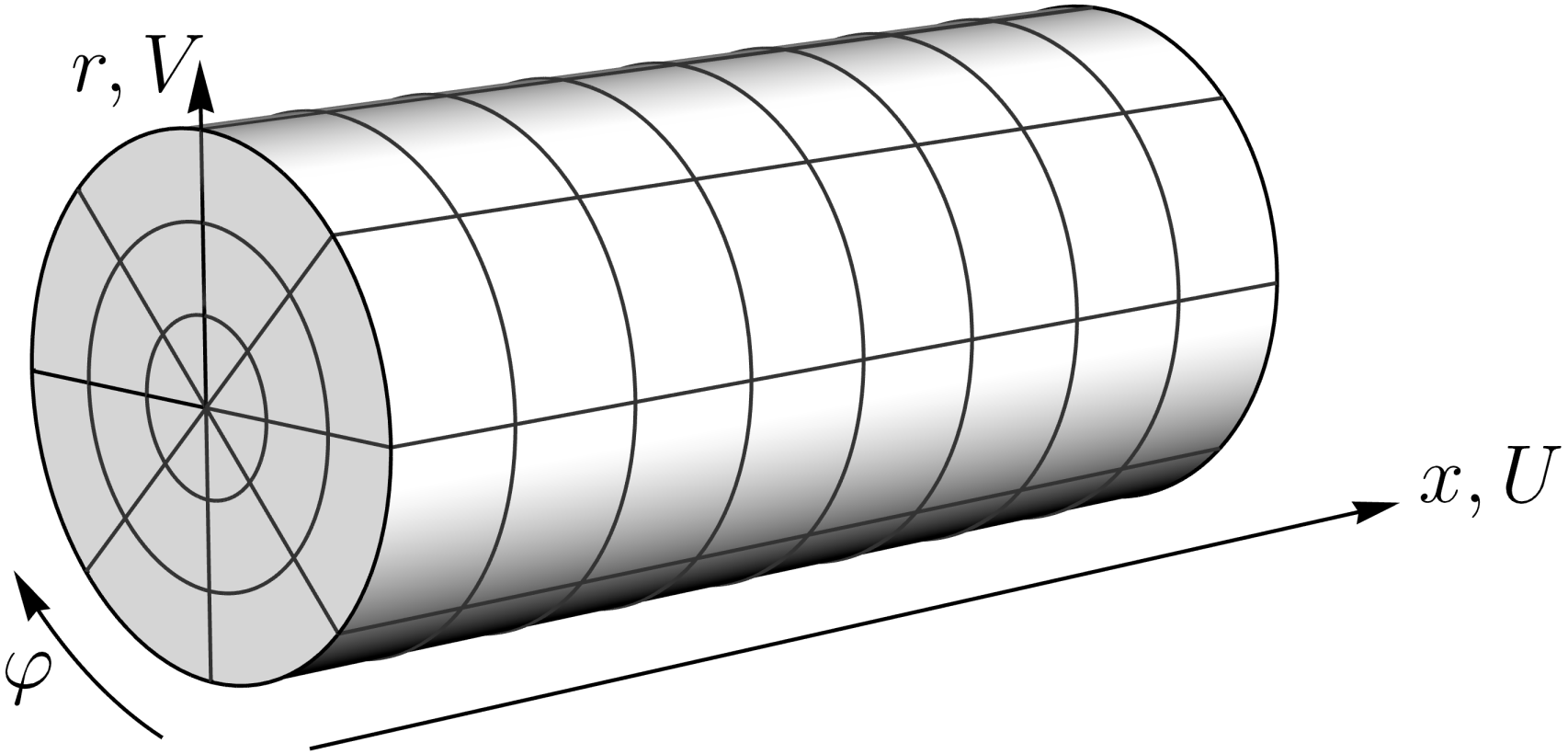}
\caption{Schematic of the circular cylindrical rod and the cylindrical coordinates.}
\label{fig:bar}
\end{figure}

Here, we describe derivation of the asymptotic model for longitudinal waves in a thin cylindrical rod shown in Fig.~\ref{fig:bar}, however the similar model can be derived for other thin waveguides, such as bars, plates and shells. 
The model derivation is based on the systematic asymptotic analysis of the full 3D equations of motion and boundary conditions in the limit of long waves, small strains and small dissipation. Here, we follow the derivation described in our previous work in Ref.~\cite{GKS2019} for the case of absolutely elastic material. We note that other approaches are also possible based on the asymptotic analysis of the equivalent variational problem formulation~\cite{SamsonovBook,Gula2018} as well as the lattice models~\cite{Khusnutdinova2009lattice}.

In the present study, we consider longitudinal axisymmetrical waves, thus rod's torsion is negligible and the displacement of rod's points is characterized by only the axial displacement~$U$ and the radial displacement~$V$. The equations of motion for the cylindrical rod take the following form:
\begin{align}
\label{eq:dyn_U}
	\rho \ddot U &= \partial_x P_{xx} + \partial_r P_{xr} 
		+ \frac{P_{xr}}{r} ,\\
\label{eq:dyn_V}
	\rho \ddot V &= \partial_x P_{rx} + \partial_r P_{rr} 
		+ \frac{P_{rr} - P_{\varphi\varphi}}{r},
\end{align}
where $P_{ij}$ denote components of the stress tensor $\tens P$,
and the free surface boundary conditions at the rod's radius $R$ are given by
\begin{align}
	P_{rr} &= 0 \quad \mbox{at} \quad r = R \label{eq:bc_rr},\\
	P_{xr} &= 0 \quad \mbox{at} \quad r = R \label{eq:bc_rx}.
\end{align}
Since the component $P_{\varphi r} \equiv 0$ due to the axial symmetry, the third boundary condition $P_{\varphi r} = 0$ at $r = R$ is always satisfied.

Axial symmetry of the problem allows application of the following power series expansions of the axial and radial displacements in the radial coordinate $r$:
\begin{align}
	\label{eq:expand_U}
	U(x,r,\varphi,t) &= U_0(x,t) + r^2 U_2(x,t) + \dots\\
	\label{eq:expand_V}
	V(x,r,\varphi,t) &= r V_1(x,t) + r^3 V_3(x,t) + \dots,
\end{align} 
Substituting these expansions into equations of motion~\eqref{eq:dyn_U} and \eqref{eq:dyn_V} and setting to zero all coefficients near different powers of the radial coordinate $r$ yields the following set of equations for the components introduced in these expansions:
\begin{align}
\label{eq:dyn_Uk}
\rho \ddot{U}_k =& \ (\lamop + \muop) \left[
	(k+2) \partial_x V_{k+1} + \partial^2_x U_k \right] 
	+ \muop \left[(k+2)^2 U_{k+2} + \partial^2_x U_k \right] 
	+ \langle \text{nonlin. terms} \rangle, \\
\label{eq:dyn_Vk}
\rho \ddot{V}_{k+1} =& \ (\lamop + \muop) \left[
	(k+2)(k+4) V_{k+3} + (k+2) \partial_x U_{k+2} \right] \nonumber\\ 
	&+ \muop \left[(k+2)(k+4) V_{k+3} + \partial^2_x V_{k+1} \right]
	+ \langle \text{nonlin. terms} \rangle, \quad k = 0, 2, 4, \dots,
\end{align}
where we do not write the nonlinear terms for brevity.

The above set of equations can be asymptotically solved for $U_{k+2}$ and $V_{k+3}$ in the limit of small strains and small dissipation. To make the analysis systematic, we introduce the following dimensionless scaled variables denoted by tilde:
\begin{gather}
	x = \frac{R_* \xs}{\delta}, \quad 
	r = \frac{R_* \rs}{\delta}, \quad 
	t = \frac{R_* \ts}{\delta c}, \quad 
	\tau_s = \frac{R_* \taus_s}{\delta c}, \nonumber\\
	U_k = \frac{\varepsilon R_*^{1-k} \Us_k}{\delta}, \quad 
	V_{k+1} = \varepsilon R_*^{-k} \Vs_{k+1}, \quad 
	\varepsilon, \delta \ll 1,
\label{eq:scales}
\end{gather}
where the small parameter $\delta$ takes into account that the characteristic wavelength $R_*/\delta$ is much larger than the gyration radius $R_*$ of the waveguide cross-section, while the small parameter $\varepsilon$ is introduced due to the small strains regime. 
For the circular cross-section of radius $R$ and the uniform density, the gyration radius writes as:
\begin{equation}\label{eq:gyr_rad}
	R_* = \frac R{\sqrt{2}}.
\end{equation}
Thus, according to the introduced scaling, the dimensionless radial coordinate $\rs$ can only take values of order $\delta$. The characteristic velocity $c$ in Eq.~\eqref{eq:scales} denotes the velocity of longitudinal infinitesimal waves in an infinitely thin rod, which has the following form:
\begin{equation}\label{eq:c}
	c^2 = \frac{E_0} \rho, \quad 
	E_0 = \frac{\mu_0(3\lambda_0 + 2\mu_0)}{\lambda_0 + \mu_0},
\end{equation}
where $E_0$ is the quasi-static Young's modulus.
In the case of small dissipation, the retarded part of the linear operators $\lamop$ and $\muop$ is assumed to be small:
\begin{equation}\label{eq:small_visc}
	\lamop[f] = \lambda_0 f + \sum_{s=1}^q \lambda_s \Rop_s[f], \quad
	\lambda_s \ll \lambda_0,
\end{equation}
and similarly $\mu_s \ll \mu_0$ for $\muop$.

Solving Eqs.~\eqref{eq:dyn_U} and \eqref{eq:dyn_V} asymptotically for $U_2$ and $V_3$ and substituting them into the first boundary condition~\eqref{eq:bc_rr} yields:
\begin{align}
	2 (\lamop + \muop) [\Vs_1] + \lamop [\partial_{\xs}\Us_0] 
	+ \varepsilon f_\text{nl}[\partial_{\xs} \Us_0, \Vs_1]
	 + \frac{\delta^2}{8} \biggl( 
		\frac{\rho c^2 \partial_{\ts}^2 \bigl(
			(4\lambda_0 + 6\mu_0) \Vs_1 
			- (\lambda_0 + 3\mu_0) \partial_{\xs} \Us_0 \bigr)}
		{\lambda_0 + 2\mu_0} \nonumber\\
	 + (\lambda_0 + 3\mu_0)\partial_{\xs}^3\Us_0 + 2\lambda_0\partial_{\xs}^2\Vs_1 
	\biggr) 
	+ \dots = 0, \label{eq:bc_rr_subst}\\
	f_\text{nl}[u, v] = (4\lop + 2\mop)[v,v] + (4\lop - 2\mop + \nop)[u,v] + \lop[u,u] + 3(\lambda_0 + \mu_0) v^2 + \lambda_0 uv + \frac{\lambda_0}2 u^2, \nonumber
\end{align}
where dots denote smaller terms of higher orders with respect to small parameters $\varepsilon$ and $\delta$.
This equation can also be solved asymptotically for $\Vs_1$ leading to the following form of the second boundary condition~\eqref{eq:bc_rx}:
\newcommand{\betaop}{\hat\beta}
\newcommand{\Es}{\tilde E}
\newcommand{\Eop}{\hat E}
\begin{equation}\label{eq:bq_damp}
\partial_{\ts}^2 \us - \partial_{\xs}^2 \biggl(
	\us + \sum_{s=1}^q \gamma_s \Rop_s[\us]	
	+ \varepsilon\sum_{s=0}^q\sum_{u=0}^q 
		\beta_{su} \Rop_s[\us] \Rop_u[\us]
	+ \delta^2 \nu_0^2 \partial_{\ts}^2{\us} + \dots
	\biggr) = 0.
\end{equation}
Here, the coefficients $\gamma_s$ and the nonlinear parameters $\beta_{su}$ are defined as follows:
\begin{align}
\label{eq:gamma_s}
	\gamma_s &= \left( 
		\frac{2(1+\nu_0)^2}{3(1+\nu_s)} + 
		\frac{(1-2\nu_0)^2}{3(1-2\nu_s)} 
	\right) \frac{E_s}{E_0},\\
\label{eq:beta_su}
	\beta_{su} &= \frac32\delta_{s0} \delta_{u0}
		+ \frac{(1 - 2\nu_0)^3 l_{su}
		+ 2(1 + \nu_0)^2 (1 - 2\nu_0)m_{su} 
		+ 3\nu_0^2 n_{su}}{E_0},
\end{align}
where $\delta_{su}$ denotes the Kronecker delta, and for convenience we introduced the viscoelastic Young's moduli $E_s$ and Poisson's ratios $\nu_s$ defined as follows:
\begin{equation}\label{eq:visc_young_poiss}
	E_s = \frac{\mu_s(3\lambda_s + 2\mu_s)}{\lambda_s + \mu_s}, \quad
	\nu_s = \frac{\lambda_s}{2(\lambda_s + \mu_s)}.
\end{equation}
Note, that Eq.~\eqref{eq:bq_damp} implies the following asymptotic relation: 
\begin{equation}\label{eq:regul}
\partial_{\ts}^2 \us = \partial_{\xs}^2 \us + \Obig(\varepsilon, \delta^2, \gamma_s),
\end{equation}
which was used to merge all the dispersive terms (fourth order derivatives) arising in the equation into the single term $\partial_{\xs}^2\partial_{\ts}^2 \us$. We also note, that if Poisson's ratio does not exhibit notable frequency dependence then all the values $\nu_s$ are equal to each other, which simplifies the form of parameters $\gamma_s$:
\begin{equation}\label{eq:gamma_s_const_nu}
	\gamma_s = \frac{E_s}{E_0}, \quad \nu_s = \nu_0, \quad
    s = 1, 2, \dots
\end{equation}

The Boussinesq-type equation~\eqref{eq:bq_damp} can be used as a model of the long longitudinal waves in nonlinear viscoelastic waveguides. In dimensional form it writes as:
\begin{equation}\label{eq:bq_damp_dim}
\frac1{c^2} \partial_{t}^2 u - \partial_x^2 \biggl(
	u + \sum_{s=1}^q \gamma_s \Rop_s[u]	
	+ \varepsilon\sum_{s=0}^q\sum_{u=0}^q 
		\beta_{su} \Rop_s[u] \Rop_u[u]
	+ \frac{R_*^2 \nu_0^2}{c^2} \partial_t^2{u} \biggr) = 0.
\end{equation}
Despite being derived for the circular cylindrical rod, this equation should also be valid in the case of other shapes of cross-section, e.g. the rectangular bar. The waveguide shape is accounted for through the gyration radius of its cross-section $R_*$, thus affecting only the dispersive coefficient, while the nonlinear coefficients remain unchanged for the waveguides of different shapes~\cite{SamsonovBook, Gula2018}.

\section{Slowly decaying solitary waves}

The Boussinesq-type equations, which have the form of the wave equation with the addition of nonlinear and dispersive terms, are known to possess the solitary wave solutions. These waves propagate with constant speed and shape due to the balance between the effects of dispersion and nonlinearity, however, if dissipation is present in the model the waves are subject to attenuation. There are two terms in Eq.~\eqref{eq:bq_damp} containing the retarded operators which are responsible for dissipation: the first one is linear and the other is nonlinear. Our further asymptotic analysis will be focused on the study of slowly decaying solitons due to the presence of the linear and nonlinear dissipation.

\subsection{Linear dissipation}
We begin the analysis from the case of small linear viscoelastic dissipation, neglecting the nonlinear dissipative effects. In this case, the linear viscoelastic moduli are small compared to the linear quasi-static moduli and all coefficients $\beta_{su}$ ecxept for $\beta_{00}$ vanish:
\begin{align}\label{eq:weak_lin_visc}
\gamma_s \ll 1, \quad s \geqslant 1 \quad \text{and} \quad
\beta_{su} = 0, \quad (s,u)\neq(0,0),
\end{align}
so that Eq.~\eqref{eq:bq_damp} after neglecting higher-order terms takes the following form:
\begin{align}
\label{eq:bq_lin_visc}
	&\partial_{\ts}^2 \us - \partial_{\xs}^2\bigg(
		\us + \sum_{s\geqslant 1}\gamma_s\Rop_s[\us]
		+ \varepsilon\beta_{00} \us^2 
		+ \delta^2 \nu_0^2 \partial_{\ts}^2 \us
		\bigg) = 0.
\end{align}
In other words, the above assumptions imply that the linear elastic moduli weakly depend on wave frequency while the nonlinear moduli are frequency independent.

Solitary waves in nonlinearly elastic and linearly viscoelastic materials were studied in other works~\cite{Nariboli1970, Zabolotskaya2004, KhusnPRE2021}, however, only the cases of high- and low-frequency dissipation were considered there. In these cases relaxation times are assumed to be either small or large compared to the characteristic time-width of a soliton, which allows derivation of asymptotic damped models without the retarded integral operators. In the preset study, we consider the more general case of arbitrary relaxation times including those of the same order as the soliton time-width.

To find the approximate decaying solitary wave solutions of Eq.~\eqref{eq:bq_lin_visc} we assume balance between the nonlinear and dispersive terms ($\varepsilon\sim\delta^2$) and smallness of dissipation ($\varepsilon^2 \ll \gamma_s \ll \varepsilon$). Then, we apply the standard method of multiple scales to derive an asymptotic model for the waves propagating in one (say, positive) direction of the $x$-axis~\cite{Johnson2005,GBK2020}. It yields the Korteweg-de Vries (KdV) equation with small dissipation, the derivation of which is presented in the Appendix. Using the theory presented in Ref.~\cite{Karpman1977}, the slowly decaying, widening, and decelerating solitary wave solution is obtained. In the dimensional form it writes
\begin{equation}\label{eq:sol_dim}
    u(x, t) = A(t) \cosh^{-2} \frac{x - x_0(t)}{L(t)}, \quad 
    A(t) = \frac{6\nu_0^2 R_*^2}{\beta_{00} L^2(t)},
\end{equation}
where the width and position of the wave are subject to the following ordinary differential equations:
\begin{align}
\label{eq:L_dim}
	L'(t) &= \frac{c}2 \sum_{s\geqslant 1} \gamma_s
	I_{1,\text{lin}}\bigg(\frac{c\tau_s}{L(t)}\bigg),\\
\label{eq:x0_dim}
	x_0'(t) &= c\biggl(1 + \frac{2\nu_0^2 R_*^2}{L(t)^2}
    + \frac12 \sum_{s\geqslant 1} \gamma_s
    I_{2,\text{lin}}\bigg(\frac{c\tau_s}{L(t)}\bigg)\biggr).
\end{align}
The introduced functions $I_{1,\text{lin}}$ and $I_{2,\text{lin}}$ define the widening and deceleration rates and take the following form:
\begin{align}
\label{eq:I1_lin}
I_{1,\text{lin}}(\theta) &= \frac{\Psi(\theta)}{2\theta^4}, \quad
\Psi(\theta) = 2\theta + 2\theta^2 
    + \frac{4\theta^3}{3} - \psi'\bigl((2\theta)^{-1}\bigr),\\
\label{eq:I2_lin}
I_{2,\text{lin}}(\theta) &= \frac{\Psi'(\theta)}{4\theta^2},
\end{align}
where $\psi(x) = \Gamma'(x)/\Gamma(x)$ is the digamma function and the prime denotes derivative. The details of the derivation are given in the Appendix.

\begin{figure}[h]
\centering
\includegraphics[width=0.5\linewidth]{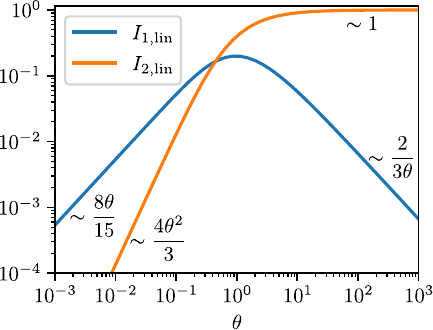}
\caption{Plots and asymptotics of the functions $I_{1,\text{lin}}(\theta)$ and $I_{2,\text{lin}}(\theta)$.}
\label{fig:plot_I1I2}
\end{figure}

Functions $I_{1,\text{lin}}$ and $I_{2,\text{lin}}$ have the following asymptotics:
\begin{align}
	&I_{1,\text{lin}}(\theta)\sim \frac{8\theta}{15},
	&&I_{2,\text{lin}}(\theta)\sim \frac{4\theta^2}3 
	&&\text{at } \theta\to0,\\
	&I_{1,\text{lin}}(\theta)\sim \frac2{3\theta},
	&&I_{2,\text{lin}}(\theta)\sim 1 
	&&\text{at } \theta\to\infty.
\end{align}
The plots of these functions together with their asymptotics are shown in Fig.~\ref{fig:plot_I1I2}.

Let us analyze the limiting cases of a long and a short solitary wave (compared to $c\tau_1$) in the case of a single relaxation time $\tau_1$. Note that in both cases soliton width is still assumed to be larger than the waveguide thickness $L\gg R_*$ since the Boussinesq-type model~\eqref{eq:bq_lin_visc} was derived under this assumption.

In the first case ($L\gg c\tau_1$) the system~\eqref{eq:L_dim} and \eqref{eq:x0_dim} can be solved exactly, resulting in the explicit expression for the time-dependent soliton width and position:
\begin{align}
    L_\text{long}(t) &= L_0 \sqrt{1 + D_1 t}, &&
    D_1 = \frac{8 \gamma_1 c^2\tau_1}{15 L_0^2}, \\
    x_{0,\text{long}}(t) &= c \left(t 
    	+ \frac{2\nu_0^2 R_*^2}{L_0^2 D_1} \ln(1 + D_1 t) \right), &&
\end{align}
where $L_0 = L(0)$ is the initial width of the soliton.
This was obtained assuming $I_1(\theta) \approx 8\theta/15$, $I_2 \approx 0$ ($\theta$ is small), and that initially the soliton was at the origin: $x_0(0) = 0$. This solution coincides with the one obtained for the case of high-frequency dissipation described by the Korteweg-de Vries-Burgers equation~\cite{Karpman1977}. Here, the soliton propagates with velocities slightly greater than the linear velocity $c$ but approaching $c$ in the process of decay.

In the second case (a short soliton, $L\ll c\tau_1$) the result is as follows:
\begin{align}
    L_\text{short}(t) &= L_0 e^{\frac{D_2 t}2}, &&
    D_2 = \frac{2\gamma_1}{3 \tau_1}, \\
    x_{0,\text{short}}(t) &= c \left(t\left(1+\frac{\gamma_1}2\right)
        + \frac{2\nu_0^2 R_*^2}{L_0^2 D_2} (1 - e^{-D_2 t})\right), &&
\end{align}
where the linear (infinitesimal strain wave) velocity is $c(1 + \gamma_1/2)$. 

The evolution of a soliton is perhaps easier to understand in terms of its signed amplitude $A(t)$, related to width $L(t)$ via Eq.~\eqref{eq:sol_dim}, and velocity $x_0'(t)$: 
\begin{align}\label{eq:decay_lin}
	A_\text{long}(t) &= \frac{A_0}{1+D_1t},  
	&&x_{0,\text{long}}'(t) = c \left(1 + \frac{A_0\beta_{00}}3
		\frac{1}{1+D_1t}\right) \\
	A_\text{short}(t) &= A_0 e^{-D_2t},  
	&&x_{0,\text{short}}'(t) = c \left(1 + \frac{\gamma_1}2 
		+ \frac{A_0\beta_{00}}3 e^{-D_2t}\right) 
\end{align}
Interestingly, the short soliton decays exponentially, while the long soliton decays as inverse time.

The difference in linear wave velocities can be intuitively explained using the spring-dashpot model. Consider the system shown in Fig.~\ref{fig:gen_sls}, with only the upper spring and a single spring-dashpot (Maxwell) element. For slow deformations (much slower than the relaxation time), the dashpot deforms fully, and the system behaves as if only the upper spring is active. Conversely, for fast deformations (much faster than the relaxation time), the dashpot almost does not deform, and the system effectively acts like two springs connected in parallel, increasing the stiffness and, therefore, the strain wave velocity.

\subsection{Nonlinear dissipation}

\begin{figure}[h]
\centering
\includegraphics[width=.9\linewidth]{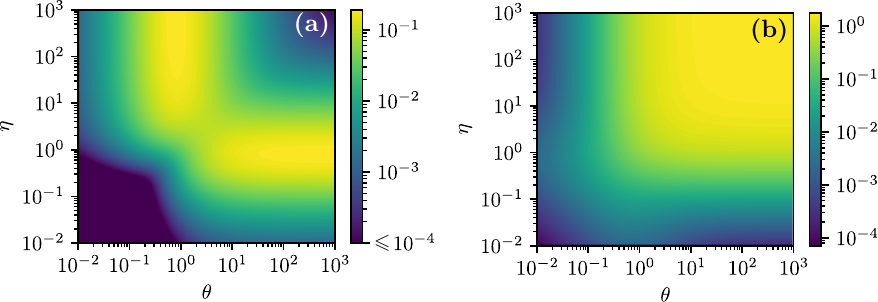}
\caption{Heatmaps of functions $I_{1,\text{nl}}(\theta, \eta)$ (panel a) and $-I_{2,\text{nl}}(\theta, \eta)$ (panel b).}
\label{fig:plot_I12nl}
\end{figure}

Here we consider the more general case of nonlinear but still relatively small dissipation. In other words, we assume that both linear and nonlinear elastic moduli are weakly dependent on wave frequency, so that instead of Eq.~\eqref{eq:weak_lin_visc} the following holds:
\begin{equation}\label{eq:weak_nonlin_visc}
	\gamma_s \ll 1, \quad \beta_{su} \ll 1, \quad (s,u)\neq(0,0).
\end{equation}
Under these assumptions all dissipative terms in Eq.~\eqref{eq:bq_damp} are small. Then, assuming $\varepsilon\sim\delta^2$ and $\varepsilon^2 \ll \gamma_s \ll \varepsilon$, and repeating the asymptotic analysis carried out in the previous section, we derive the more general equations compared to Eqs.~\eqref{eq:L_dim} and \eqref{eq:x0_dim}, which govern the decay of the soliton in Eq.~\eqref{eq:sol_dim}:
\begin{align}
\label{eq:L_nl_dim}
	L'(t) &= c \biggl( \frac12\sum_{s\geqslant 1} \gamma_s
		I_{1,\text{lin}}\bigg(\frac{c\tau_s}{L}\bigg)
		+ \frac{\nu_0^2 R_*^2}{2L^2} \mathop{\sum\sum}_{(s,u)\neq(0,0)} 
		\beta_{su} I_{1,\text{nl}} 
		\biggl(\frac{c\tau_s}L,\frac{c\tau_u}L\biggr) \biggr),\\
\label{eq:x0_nl_dim}
	x_0'(t) &= c\biggl(1 + \frac{2\nu_0^2 R_*^2}{L^2}
	 + \frac12 \sum_{s\geqslant 1} \gamma_s
	 I_{2,\text{lin}}\bigg(\frac{c\tau_s}L\bigg) 
	 + \frac{\nu_0^2 R_*^2}{2L^2} \mathop{\sum\sum}_{(s,u)\neq(0,0)} 
	 \beta_{su} I_{2,\text{nl}} 
	 \biggl(\frac{c\tau_s}L,\frac{c\tau_u}L\biggr) \biggl).
\end{align}
The introduced functions $I_{1,\text{nl}}$ and $I_{2,\text{nl}}$ define the nonlinear corrections to the widening and deceleration rates of the soliton and take the form:
\begin{align}
\label{eq:I1_nl}
	I_{1,\text{nl}}(\theta,\eta) &= 
	2\int\limits_{-\infty}^{+\infty} \cosh^{-2}z \, \tanh z \,
		J(\theta,z) J(\eta,z) \dd{z},\\
\label{eq:I2_nl}
	I_{2,\text{nl}}(\theta,\eta) &= 2\int\limits_{-\infty}^{+\infty} 
		\left(z \tanh z - 1 \right) \cosh^{-2}z \,
		J(\theta,z) J(\eta,z) \dd{z},\\
\label{eq:J}
	J(\theta,z) &= \frac1\theta\left( 1 - \tanh z 
		-\frac{2e^{-2z} {_2F_1(1, 1+\frac1{2\theta}, 
			2+\frac1{2\theta}; -e^{-2z})}
		}{1+2\theta} \right) - \cosh^{-2}z.
\end{align}
Unfortunately, we did not find a simple representation of these functions even with the help of special functions. The plots of these functions are shown in Fig.~\ref{fig:plot_I12nl}.

\section{Comparison with 3D simulations}

In this section we compare the derived theory of solitary wave decay with the results of the direct 3D simulation of the equations of motion~\eqref{eq:motion} with the boundary conditions~\eqref{eq:bc}. 

The linear viscoelastic moduli and density of the rod's material used in the simulations are listed in Tab.~\ref{tab:lin_mod}. Consistent with our previous studies on waves in glassy polymers, we set the relaxation times to be equidistantly distributed in logarithmic space, with the same linear viscoelastic moduli assigned to each relaxation time. However, the values of the linear viscoelastic moduli were chosen to be significantly smaller than those typically observed in real glassy polymers~\cite{WaMot2022}. This choice ensures that the model is tested within its range of validity, requiring the ratio $\gamma_s= E_s / E_0$ to be much smaller than the characteristic strain.
\begin{table}[h]
    \centering
    \begin{tabular}{|l|l|l|}
        \hline
        \textbf{Parameter} & \textbf{Symbol} & \textbf{Value} \\
        \hline
        Density           & $\rho$          & $\text{1}~\text{mg/mm}^3$ \\
        Radius			  & $R$ 			& 1 mm \\
        Gyration radius   & $R_*$			& 0.7 mm \\
        Poisson's Ratio   & $\nu_0$ & 0.33 \\
        Quasistatic Young's modulus   & $E_0$             & 4~GPa \\
        Viscoelastic Young's moduli & $E_1, \dots E_7$        & $1.3\cdot 10^{-3}$~GPa \\
        Linear viscoelastic parameters & $\gamma_1, \dots \gamma_7$        & $0.33\cdot 10^{-3}$ \\
        Relaxation times    & $\tau_1, \dots \tau_7$      & 0.1, 0.3, 1, 3, 10, 30, 100~$\mu$s \\
        \hline
    \end{tabular}
    \caption{Density, geometry of the rod's cross-section and linear viscoelastic parameters of the rod's material used in simulations.}
    \label{tab:lin_mod}
\end{table}

The nonlinear parameters are defined by the three symmetric $8\times 8$ matrices $\{l_{su}\}$, $\{m_{su}\}$, and $\{n_{su}\}$, each corresponding to one of the nonlinear Murnaghan moduli. The size of these matrices is determined by number of relaxation times: there are 7 relaxation times listed in Tab.~\ref{tab:lin_mod} and the additional infinite $\tau_0 = \infty$ is used to account for quasistatic elastic properties. Each component of these matrices, say $l_{su}$, is related to two relaxation times $\tau_s$ and $\tau_u$. 
We assume that the components of these matrices are nonzero only for $\tau_0 = \infty$ and $\tau_3 = 1\ \mu s$, since this was shown to be enough to describe experimentally measured frequency-dependent Murnaghan moduli~\cite{IJNLM2024}. Thus, each matrix has only 4 nonzero components, which are listed in Tab.~\ref{tab:nonlin_mod}.
\begin{table}[h]
    \centering
    \begin{tabular}{|l|l|l|}
        \hline
        \textbf{Parameter} & \textbf{Symbol} & \textbf{Value} \\
        \hline
        Murnaghan modulus $l$   & $l_{00},\ l_{03} = l_{30},\ l_{33}$             & $-20,\ 1,\ -1~\text{GPa}$ \\
        Murnaghan modulus $m$   & $m_{00},\ m_{03} = m_{30},\ m_{33}$             & $-15,\ 1,\ -1~\text{GPa}$ \\
        Murnaghan modulus $n$   & $n_{00},\ n_{03} = n_{30},\ n_{33}$             & $-10,\ 1,\ -1~\text{GPa}$ \\
        Nonlinear viscoelastic parameters  & $\beta_{00},\ \beta_{03} = \beta_{30},\ \beta_{33}$     & $-5.5,\ 0.4,\ -0.4$ \\
        \hline
    \end{tabular}
    \caption{Nonlinear viscoelastic parameters of the model material used in simulations.}
    \label{tab:nonlin_mod}
\end{table}

In 3D simulations, the initial condition for the longitudinal displacement $U$ was set such that the longitudinal strain remained uniform across the cross-section.
This strain profile was given the solitary wave shape defined in Eq.~\eqref{eq:sol_dim} with the amplitude $2\cdot 10^{-3}$:
\begin{gather}
\label{eq:sol_dim_sim}
    \partial_x U(x,r,\varphi,0) = A_0 \cosh^{-2} \frac{x}{L_0}, \quad 
    L_0 =  \frac{R_* |\nu_0|}{\sqrt{A_0\beta_{00} / 6}}, \quad
    A_0 = -2\cdot 10^{-3}.
\end{gather}
The radial displacement was determined based on the leading-order term in expansion~\eqref{eq:expand_V}, commonly referred to as Love's hypothesis~\cite{Saccomandi2024}, which can be derived from Eqs.~\eqref{eq:small_visc} and \eqref{eq:bc_rr_subst}:
\begin{gather}
\label{eq:plane_sec_sim}
    V(x,r,\varphi,0) = -\nu_0 r \partial_x U(x,r,\varphi,0),
\end{gather}
and the internal variables $\tens E_1, \dots \tens E_7$ were set to zero. 

The rod was modeled with a radius of 1 mm and a length of 300 mm, with periodic boundary conditions imposed along its longitudinal edges, which allowed the soliton to propagate indefinitely.
To ensure that any waves emitted by the soliton did not interfere with it, a sponge layer was added at the ends of the 300 mm window, moving with the linear wave velocity $c$.

For spatial discretization, we used the multidomain pseudospectral method used in our previous works~\cite{GBK2020, WaMot2022} and described in~\cite{ICNAAM2021}. The mesh consisted of 300 domains along the rod's axis ($x$), with each domain containing 10 points in both the axial ($x$) and radial ($r$) directions. Due to axial symmetry, the angular axis ($\varphi$) was discretized with a single point. For temporal discretization, we applied Runge-Kutta 8(5,3) method with adaptive time step.

\begin{figure}[h]
	\centering
	\includegraphics[width=.9\linewidth]{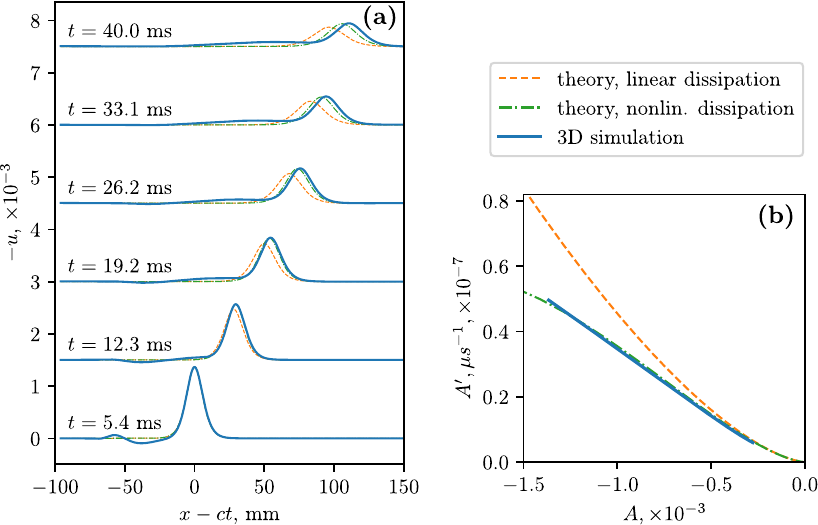}
	\caption{Comparison of 3D simulation results with the theoretical models of solitary wave decay. (a) Wave profiles at different time moments, plotted as functions of the moving coordinate. To differentiate the curves, the vertical shift by 1.5 is applied. The 3D simulation curve represents the negative linear strain at the rod's axis, $-{\partial U(x,0,0)}/{\partial x}$. (b) Phase portrait of Eqs.~\eqref{eq:L_dim} and \eqref{eq:L_nl_dim} governing the soliton decay rate expressed in terms of the amplitude parameter $A(t)$ (related to $L(t)$ via Eq.\eqref{eq:sol_dim}) compared to the phase portrait obtained from the 3D simulation.}
	\label{fig:plot_3d}
\end{figure}

Comparison of the 3D simulation with the theory, derived in the previous section, is shown in Fig.~\ref{fig:plot_3d}. Since the initial conditions in Eqs.~\eqref{eq:sol_dim_sim} -- \eqref{eq:plane_sec_sim} for the 3D simulations describe the solitary wave only approximately, it takes some time for the initial wave to evolve into the slowly decaying soliton, emitting radiation in the process. For this reason, in Fig.~\ref{fig:plot_3d}(a) the results are shown for $t \geqslant 5.4$~ms when the soliton becomes clearly separated from the oscillations radiated by the initial wave.

Figure~\ref{fig:plot_3d}a shows the negative linear strain at the rod's axis obtained in the 3D simulations, together with the theoretical decaying solitary wave solutions obtained by integrating Eqs.~\eqref{eq:L_dim} and \eqref{eq:x0_dim} for the linear theory and Eqs.~\eqref{eq:L_nl_dim} and \eqref{eq:x0_nl_dim} for the nonlinear theory. 
Figure~\ref{fig:plot_3d}b displays the phase portrait of equations governing the soliton decay rate (Eqs.~\eqref{eq:L_dim} and \eqref{eq:L_nl_dim}) in terms of amplitude parameter $A(t)$ instead of the width parameter $L(t)$. The relationship between $A(t)$ and $L(t)$ is established in Eq.~\eqref{eq:sol_dim}.
It is evident that the nonlinear theory demonstrates better agreement with the 3D simulation, particularly regarding the relationship between $A'(t)$ and $A(t)$. We could have also shown the dependence of $x_0'(t)$ on $A(t)$ to check the applicability of Eqs.~\eqref{eq:x0_dim} and \eqref{eq:x0_nl_dim}. However, the curves corresponding to the linear and nonlinear dissipation theories, as well as the 3D simulations, would be nearly indistinguishable on such a plot. Therefore, we have opted not to include it.

The developed theory can be extended further to describe also the tail behind the soliton~\cite{OstrovskyBook2022} in addition to the soliton itself, however, we leave it for future work.

Several published works have provided estimates for the linear~\cite{WaMot2022} and nonlinear~\cite{IJNLM2024, Belashov2024} viscoelastic parameters of polystyrene. These estimates differ significantly from the values listed in Tables~\ref{tab:lin_mod} and~\ref{tab:nonlin_mod}, particularly in the viscous Young's moduli $E_1, E_2, \dots$, which are two orders of magnitude larger. Similarly, the nonlinear moduli $l_{su}$, $m_{su}$, and $n_{su}$ are much larger in these works. 
These higher values exceed the limits required for the assumptions of sufficiently small dissipation (Eq.~\eqref{eq:weak_nonlin_visc}) to hold, at least for reasonable strain soliton amplitudes ($\lesssim 10^{-3}$). Nevertheless, we tested the developed theory in this context. Two cases were considered: the first involved increasing the linear viscous moduli while keeping the nonlinear moduli as given in Tab.~\ref{tab:nonlin_mod}, and the second involved increasing both the linear and nonlinear moduli. 
In the first case, the linear moduli $E_1, \dots E_7$  were scaled up by a factor of 100 compared to the values in Tab.~\ref{tab:lin_mod}, while the increased nonlinear moduli for the second case are listed in Tab.~\ref{tab:nonlin_mod2}.
\begin{table}[h]
	\centering
	\begin{tabular}{|l|l|l|}
		\hline
		\textbf{Parameter} & \textbf{Symbol} & \textbf{Value} \\
		\hline
		Murnaghan modulus $l$   & $l_{00},\ l_{03} = l_{30},\ l_{33}$             & $-800,\ 780,\ -780~\text{GPa}$ \\
		Murnaghan modulus $m$   & $m_{00},\ m_{03} = m_{30},\ m_{33}$             & $-600,\ 585,\ -585~\text{GPa}$ \\
		Murnaghan modulus $n$   & $n_{00},\ n_{03} = n_{30},\ n_{33}$             & $-400,\ 390,\ -390~\text{GPa}$ \\
		Nonlinear viscoelastic parameters  & $\beta_{00},\ \beta_{03} = \beta_{30},\ \beta_{33}$     & $-219,\ 215,\ -215$ \\
		\hline
	\end{tabular}
	\caption{Increased nonlinear viscoelastic parameters of the material used in simulations.}
	\label{tab:nonlin_mod2}
\end{table}

\begin{figure}[h]
	\centering
	\includegraphics[width=\linewidth]{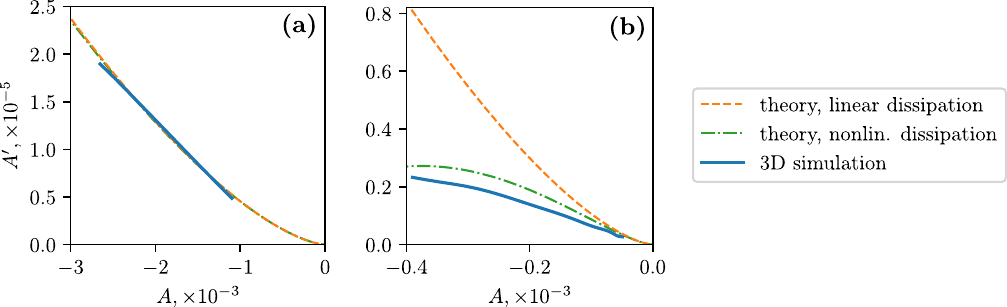}
	\caption{Phase portraits of soliton decay in rods with different material properties. (a)~Material with parameters from Tabs.~\ref{tab:lin_mod} and \ref{tab:nonlin_mod}, but with linear viscous moduli $E_1, \dots, E_7$ increased by a factor of 100. (b)~Same material as in (a), but with nonlinear moduli replaced by those listed in Tab.~\ref{tab:nonlin_mod2}.}
	\label{fig:mat34}
\end{figure}

The results are presented in Fig.~\ref{fig:mat34}. For the case of increased linear viscous moduli (Fig.~\ref{fig:mat34}a), the predictions of the linear and nonlinear theories are nearly identical within the considered range of amplitudes and closely match the 3D numerical simulation results. 
When both the linear and nonlinear moduli are increased (Fig.~\ref{fig:mat34}b), both theories show notable discrepancies from the 3D simulation results. 
Nevertheless, the nonlinear theory produces the phase portrait that is qualitatively similar to the one obtained from the 3D simulation and offers a significantly more accurate quantitative description compared to the linear theory.

\begin{figure}[h]
	\centering
	\includegraphics[width=.7\linewidth]{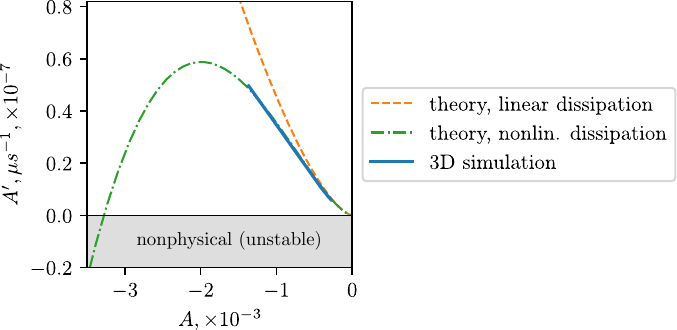}
	\caption{Same as in Fig.~\ref{fig:plot_3d}b, but in wider amplitude range.}
	\label{fig:phase_large_ampl}
\end{figure}

Let us now address the limitations of the developed model of nonlinear dissipation. A notable drawback of this theory is that it yields nonphysical results at sufficiently large soliton amplitudes, manifesting as negative dissipation (see Fig.~\ref{fig:phase_large_ampl}). The critical point, where the dissipation in nonlinear model vanishes, corresponds to the non-zero steady-state solution of Eq.~\eqref{eq:L_nl_dim}, and at larger amplitudes the soliton becomes unstable.
These limitations are similar to those associated with the Murnaghan elastic model, which assumes a cubic expression for the specific elastic energy. In both cases, the assumptions inherent to the model lead to instability at sufficiently large strain amplitudes, highlighting the need for caution when applying these models.

\section{Conclusion}

In this paper, we systematically developed the model for the longitudinal waves in thin rods and bars made of nonlinear viscoelastic materials (damped Boussinesq-type equation~\eqref{eq:bq_damp_dim}). The linear dissipation is modeled using linear retarded operators acting on the longitudinal strain, while the nonlinear dissipation is represented as similar bilinear operators.

The solitary waves governed by the derived equation, propagating in one direction, can be approximated by the damped KdV equation. By assuming small dissipation, we treated this equation as nearly integrable, allowing us to apply established theoretical methods to derive the decaying solitary wave solution (Eqs.~\eqref{eq:sol_dim}, \eqref{eq:L_nl_dim}, and \eqref{eq:x0_nl_dim}).

The developed theory of decaying solitons was compared against the solitary wave propagation data obtained in 3D simulations. The inclusion of nonlinear dissipation in the theoretical model showed excellent agreement with the 3D simulations (Fig.~\ref{fig:plot_3d}), validating the proposed approach. We showed that the theory can give a reasonable description of soliton dissipation even when the values of viscoelastic moduli do not satisfy the assumptions of the model (Fig.~\ref{fig:mat34}).

We have demonstrated that the frequency dependence of the third-order elastic constants (Murnaghan moduli) has a significant impact on the soliton decay rate. Specifically, for viscoelastic moduli similar to those estimated for polystyrene, the decay rate is several times smaller than predicted by the linear dissipation model (Fig.~\ref{fig:mat34}b). This finding is particularly important for potential applications of solitons, as it highlights the need to account for nonlinear dissipation effects to achieve more accurate predictions.

\section*{Acknowledgements}
This work was supported by the Russian Science Foundation, project no.~22-72-10083,\\
https://www.rscf.ru/en/project/22-72-10083/.

\appendix
\section{Derivation of the decaying solitary wave solution}

\subsection{Linear dissipation}
Assume the function $\us$ to have the following form:
\begin{equation}\label{eq:multi_scale}
\us(\xs,\ts) = \us(\xi, T), \quad \xi = \xs - \ts, \quad T = \varepsilon\ts,
\end{equation}
which describes waves propagating along the characteristic $\xi$, and the slow-time variable $\tau$ allows to capture the slow deviation of this wave from the D'Alambert solution of the simple wave equation.
With this assumption, Eq.~\eqref{eq:bq_lin_visc} is satisfied in the leading order in $\varepsilon$. In the next order it takes the form of the Korteweg--de Vries equation with dissipation:
\begin{equation} \label{eq:kdv_lin_visc}
\partial_T \us + \beta_{00} \us \partial_{\xi} \us
	+ \frac{q}2 \partial_{\xi}^3 \us 
	= \sum_{s\geqslant 1} \frac{\gamma_s}{2\varepsilon} 
		\partial_{\xi} \Rop_s[\us],
\end{equation} 
where we remind that $\gamma_s \ll \varepsilon$ and the retarded operator in the introduced coordinates 
takes the following form:
\begin{equation} \label{eq:E_op_xi}
\Rop_s[\us] = \int\limits_{\xi}^{+\infty}
	\exp\left({\frac{\xi-\xi_1}{\taus_s}} \right)
	\partial_{\xi_1} \us(\xi_1, T) \dd{\xi_1}.
\end{equation}

To use the results of the adiabatic approximation for solitons presented in Ref.~\cite{Karpman1977}, we transform the KdV part of Eq.~\eqref{eq:kdv_lin_visc} to the standard form:
\begin{gather}
\label{eq:kdv_scaling}
	\us = -\frac6{\beta_{00}} \ub(\xib, \breve T), \quad
	\xi = \sqrt{\frac q 2 }\xib, \quad
	T = \sqrt{\frac q 2 }\breve T, \quad
	\taus_s = \sqrt{\frac q 2 }\taub_s, \quad
	\gamma_s = 2\varepsilon\breve\gamma_s,\\
\label{eq:kdv_lin_visc_std}
	\partial_{\breve T} \ub - 6 \ub \partial_{\xib} \ub
	+ \partial_{\xib}^3 \ub = \sum_{s\geqslant 1} \breve\gamma_s 
		\partial_{\xib} \Rop_s[\ub].
\end{gather}
The solitary wave solution can be expressed as follows:
\begin{equation}\label{eq:sol_std}
\ub(\xib, \breve T) = - \frac2{\breve L^2(\breve T)} 
    \cosh^{-2} z, \quad 
z = \frac{\xib - \xib_0(\breve T)}{\breve L(\breve T)},
\end{equation}
where the width $\breve L$ and position $\xib_0$ are subject to the following system of ordinary differential equations:
\begin{align}
\label{eq:L_std_int}
	\breve L' &= \frac{\breve L^3}4 \int_{-\infty}^{\infty}
        \hat R[\ub](z) \cosh^{-2}z \dd{z},\\
\label{eq:xi0_std_int}
	\xib_0' &= \frac4{\breve L^2} - \frac{\breve L^3}4
        \int_{-\infty}^{\infty} \hat R[\ub](z)
        \left(z + \frac{\sinh 2z}2 \right) \cosh^{-2}z \dd{z},
\end{align}
and $\hat R[\ub]$ denotes the total operator in the right-hand side of the damped KdV equation~\eqref{eq:kdv_lin_visc_std} acting on the function in Eq.~\eqref{eq:sol_std}. As a function of the introduced coordinate $z$ in Eq.~\eqref{eq:sol_std}, it takes the following form:
\begin{align}
\hat R[\ub](z) &= -\frac2{\breve L^3} \sum_{s\geqslant 1} 
	\breve\gamma_s
    \partial_z \int_z^{+\infty} 
        \exp\biggl(\frac{(z-z_1) \breve L}{\taub_s}\biggr)
        \partial_{z_1} \cosh^{-2}(z_1) \dd{z_1}. \nonumber\\
	&= -\frac2{\breve L^3} \sum_{s\geqslant 1} \breve\gamma_s
		\partial_z J\biggl(\frac{\taub_s}{\breve L}, z\biggr),
\label{eq:kdv_lin_std_rhs}
\end{align}
where the expression for function $J(\theta,z)$ is written in the main text in Eq.~\eqref{eq:J}.
Substitution of the function in Eq.~\eqref{eq:kdv_lin_std_rhs} into Eqs.~\eqref{eq:L_std_int} and \eqref{eq:xi0_std_int} leads to the following system:
\begin{align}
\label{eq:L_std}
	\breve L' &= \sum_s \breve\gamma_s
        I_{1,\text{lin}}\bigg(\frac{\taub_s}{\breve L}\bigg),\\
\label{eq:xi0_std}
	\xib_0' &= \frac4{\breve L^2} + \sum_s \breve\gamma_s 
        I_{2,\text{lin}}\bigg(\frac{\taub_s}{\breve L}\bigg),
\end{align}
where the functions $I_{1,\text{lin}}$ and $I_{2,\text{lin}}$ are given in the main text (Eqs.~\eqref{eq:I1_lin} and \eqref{eq:I2_lin}).

To derive the dimensional form of the soliton solution presented in the main text, we have to descale it according to Eq.~\eqref{eq:kdv_scaling}:
\begin{align}
\label{eq:kdv_sol}
    \us(\xi, T) &= \As(T) \cosh^{-2}\frac{\xi - \xi_0(T)}{\tilde L(T)}, 
    \quad \tilde A(T) = \frac{6q}{\beta_{00} \tilde L^2(T)},\\
\label{eq:L}
	\tilde L' &= \sum_{s\geqslant 1} \frac{\gamma_s}{2\varepsilon}
    I_1\bigg(\frac{\taus_s}{\tilde L}\bigg),\\
\label{eq:xi0}
	\xi_0' &= \frac{2q}{\tilde L^2} + \sum_{s\geqslant 1} 
    \frac{\gamma_s}{2\varepsilon} I_2\bigg(\frac{\taus_s}{\tilde L}\bigg).
\end{align}
Finally, the dimensional form in Eqs.~\eqref{eq:sol_dim} -- \eqref{eq:x0_dim} can be obtained by descaling the above equations once again using the scales in Eq.~\eqref{eq:scales}.

\subsection{Nonlinear dissipation}
In the case of nonlinear dissipation the KdV equation takes the following form:
\begin{equation} \label{eq:kdv_nonlin_visc}
	\partial_T \us + \beta_{00} \us \partial_{\xi} \us
		+ \frac{q}2 \partial_{\xi}^3 \us 
		= \sum_{s\geqslant 1} \frac{\gamma_s}{2\varepsilon} 
			\partial_{\xi} \Rop_s[\us]
	+ \mathop{\sum\sum}_{(s,u)\neq(0,0)} 
	\frac{\beta_{su}}2 	\partial_{\xi} \bigl(
		\Rop_s[\us] \Rop_u[\us]\bigr).
\end{equation} 
Then we repeat the same steps as in the case of linear dissipation. First, we transform the KdV part of this equation to the standard form:
\begin{equation} \label{eq:kdv_nonlin_visc_std}
	\partial_{\breve T} \ub - 6 \ub \partial_{\xib} \ub
		+ \partial_{\xib}^3 \ub = \sum_{s\geqslant 1} \breve\gamma_s 
			\partial_{\xib} \Rop_s[\ub] 
	+ \mathop{\sum\sum}_{(s,u)\neq(0,0)} 
	\breve\beta_{su} \partial_{\xib} \bigl(
		\Rop_s[\ub] \Rop_u[\ub]\bigr),
	\quad \breve\beta_{su} = \frac{3\beta_{su}}{\beta_{00}},
\end{equation}
using the scales in Eq.~\eqref{eq:kdv_scaling}. Then, we consider the r.h.s. of this equation for the solitary wave solution in Eq.~\eqref{eq:sol_std}:
\begin{equation}\label{eq:kdv_nonlin_std_rhs}
	\hat R[\ub](z) = -\frac2{\breve L^3} \sum_{s\geqslant 1} 
		\breve\gamma_s
		\partial_z J\biggl(\frac{\taub_s}{\breve L}, z\biggr) 
	+ \frac4{\breve L^5} \mathop{\sum\sum}_{(s,u)\neq(0,0)} 
		\breve\beta_{s} \partial_z \biggl(
			J\biggl(\frac{\taub_s}{\breve L}, z\biggr)
			J\biggl(\frac{\taub_u}{\breve L}, z\biggr)
		\biggr).
\end{equation}
Finally, substitution of this function into Eqs.~\eqref{eq:L_std_int} and \eqref{eq:xi0_std_int} leads to the system in Eqs.~\eqref{eq:L_nl_dim} and \eqref{eq:x0_nl_dim} written in the main text in dimensional form.

\bibliographystyle{elsarticle-num}
\bibliography{refs_ijnlm2024}

\end{document}